\documentclass[prl,twocolumn,twocolumngrid]{revtex4}
\newcommand{\witheps}[1]{}

\usepackage{graphicx}
\usepackage{amsfonts}
\usepackage{amsmath}
\usepackage{bm}
\usepackage{alltt}
\usepackage{fancyhdr}

\begin{document}

\title{Simultaneous Oxygen and Boron Trifluoride Functionalization of
Hexagonal Boron Nitride: A Designer Cathode Material for Energy Storage
}


\author{K\'aroly N\'emeth$^{\ast}$}
\affiliation{Physics Department, Illinois Institute of Technology, Chicago, Illinois 60616,
USA \\ \email{nemeth@agni.phys.iit.edu} }


\begin{abstract}
Covalent functionalization 
is a way to tune the electrochemical properties of hexagonal boron nitride (h-BN)
monolayers.    
The wide band gap insulator h-BN may become metallic conductor upon
functionalization with strong oxidants, such as fluorosulfonyl radicals 
($\cdot$OSO$_{2}$F), as known
since 1978 [N. Bartlett et al., J. Chem. Soc. Chem. Comm. {\bf 5}, 200 (1978)],
with
electrical conductivity of 1.5 S/cm [C. Shen et al., J. Solid State
Chem. {\bf 147}, 74 (1999)] that greatly surpasses 
commercial cathode material Li$_{x}$CoO$_{2}$ while retaining excellent ionic
conductivity.
Functionalized boron nitrides (FBN-s) have great potential for 
cathode applications in energy storage devices, for example in solid state batteries.
While fluorosulfonyl functionalization is unlikely to result in rechargeable cathodes,
similarly to graphene fluoride (CF$_{x}$),
some other FBN-s discussed here may do. In the present work,  
fluorene, oxygen and combined oxygen and boron trifluoride
functionalizations are studied, on the basis of band structure
calculations. Due to the open surfaces of FBN-s, fast ionic diffusion with Li, Na
and Mg ions is possible, enabling batteries 
with voltages of 2.1-5.6 V, theoretical energy
densities of 800-1200 Wh/kg and fast charge and discharge. 
\end{abstract}


\maketitle

\section{Introduction}
Ideal electrochemical energy storage devices would simultaneously fulfill several
desirable properties, such as 
high gravimetric and volumetric energy and power densities, safety of operation, economic
and environmentally friendly composition, rechargeability, fast charging and discharging, 
and a large number of charging/discharging cycles (long cycle life) 
\cite{MArmand08,simon2008materials}.

While most traditional electroactive materials consist of three-dimensional (3D)
structures, such as crystals or amorphous materials, the recent emergence of 2D
materials offers new and advantageous platforms for novel ways of energy storage
\cite{geim2007rise,wang2016hybrid,pakdel2014nano,weng2016functionalized}.
 
The recent realization of hybrid supercapacitor-batteries through the application of
graphene oxide (GO) cathodes and Li or Na anodes demonstrated simultanous achievement
of high energy and exceptionally high power densities \cite{liu2014lithium,kim2014novel}.
Power densities of 4-45 kW/kg (on average) and specific energies of
100-500 Wh/kg have been achieved and stable capacity (300-450 mAh/g) for over 1000 cycles
of charge and discharge demonstrated, with Li anode and GO cathode, 
where the GO cathode had as much as 21-32
w\% oxygen content \cite{liu2014lithium}. With Na anode and GO cathode, energy
density was at 100-500 Wh/kg, power density at 55 kWh/kg and stable capacity
demonstrated through 300 cycles \cite{kim2014novel}. The respective cell reactions during
discharge involve the settling of one negative charge on the oxygen atoms of GO during
the ring opening of surface-bound epoxy bonds and the conversion of edge-bound
C=O (oxo) groups to phenolate-type ones \cite{kim2014novel}.

In functionalized monolayer 2D materials (F2D-s), cations can hop directly from the electrolyte to
the reactive/intercalating surface sites of F2D-s, allowing for very high power
densities (charge/discharge rates), similar to those in supercapacitors. This is not
possible in 3D materials in which diffusion of ions is normally much slower.
The advantage of 2D materials in designing high power density energy storage devices
appears to be unparalleled among 3D materials for this simple geometric
consideration. 

There are many layered materials, e.g., MoS$_{2}$, WS$_{2}$,
MoSe$_{2}$, WSe$_{2}$, and TiS$_{2}$, that can be exfoliated to monolayers. However,
in contrast to graphene (G) and h-BN, these transition metal containing
monolayers cannot offer high specific energy and power densities because of the
presence of heavy transition metal elements (per weight properties become disadvantageous). 
Furthermore, the transition metals
involved are often rare, expensive and environmentally unfriendly. Therefore, only C, N
and B based 2D monolayers appear to be practical for industrial-scale energy storage
technologies among the 2D materials.

Unfortunately, graphene oxide is thermally unstable, it easily and often
explosively decomposes to CO, CO$_{2}$ and carbon when heated to 200-300 $^{o}$C 
\cite{kim2010self,krishnan2012energetic}. This is one of the reasons why the
scalable production of GO is still in its infancy. Other than oxygen
functionalization of graphene, such as -COOH, -OH, -NH$_{2}$, -OR, and –COOR ones, have 
also been considered for cathode applications, however not tested yet
\cite{liu2014lithium}. Another problem with graphene is that it turns from an
electrically conductive material to an insulator when densely functionalized (such
as in highly oxidized graphene). As opposed to graphene, h-BN may become a small or zero
bandgap system when densely functionalized \cite{bhattacharya2012band} also
illustrated by the BN(F)-BN(OBF$_{2}$) example and its discharged (lithiated, sodiated, etc)
 versions below.
The most drastic band gap reduction has been observed
in fluorinated or fluorosulfonated h-BN experimentally, 
resulting in good electrical conductivity 
\cite{xue2013excellent,du2014one,bartlett1978novel,shen1999intercalation}.
Thus, certain functionalized h-BN-s may simultaneously be good electric and ionic 
conductors in both charged (lithiated, cathionated) and discharged (delithiated,
de-cathionated) states.

As opposed to GO, oxidized h-BN is stable even abovove 800 $^{o}$C 
\cite{li2014strong,cui2014large}. Both edge and surface functionalizations of h-BN 
have been carried out with various functional groups using radical species, such as
$\cdot$OSO$_{2}$F (fluorosulfonyl, as early as 
in 1978 \cite{bartlett1978novel,shen1999intercalation}),
$\cdot$OH \cite{RU2478077C2,nazarov2012functionalization,sainsbury2012oxygen,lin2012method,pakdel2014plasma,nayak2014inversion}, 
$\cdot$NH$_{2}$ \cite{RU2478077C2,nazarov2012functionalization,ikuno2007amine,lin2012method} ,
$:$CBr$_{2}$ \cite{sainsbury2014dibromocarbene}, 
$\cdot$F \cite{xue2013excellent,du2014one}, and substituted phenyl \cite{rajendran2012surface} functional groups.
Radical species preferencially bind to the electron-deficient B atoms, except double
radicals (such as $\cdot$O$\cdot$, $:$CBr$_{2}$) which bind to both B and N
simultaneously \cite{sainsbury2014dibromocarbene,anota2011lda}.
Covalent functionalization of h-BN has also been carried out using ionic species,
such as OH$^{-}$ \cite{lee2015scalable,Fu2016microwave,zhang2016experimental},  
NO$_{2}^{+}$ or SO$_{3}^{+}$ which bind to positively charged B and negatively
charged N atoms of h-BN \cite{RU2478077C2} and may be delivered in solution or melt
phases. Furthermore, it has been known for decades from the study of refractory
material h-BN, that the melt of certain salts, such as LiOH, KOH, NaOH-Na$_{2}$CO$_{3}$,
NaNO$_{3}$ and Li$_{3}$N etches h-BN \cite{kumashiro2000electric}, whereby
functionalized h-BN species form \cite{yamane1987high}.
The surface coverage of h-BN by -OH and -NH$_{2}$ functional groups 
may be as high as about 30 atom\%
as shown by XPS analysis \cite{nazarov2012functionalization}.
The reactions between functionalizing radicals or ions and the h-BN surface are
based on Lewis acid/base reactions, utilizing the Lewis acid character of the B
and Lewis base character of N atoms in h-BN that can react with strong Lewis
bases/acids, respectively.
A recent review by Bando, Golberg and colleagues discusses the advances in
functionalization of h-BN in further details \cite{weng2016functionalized}.

The high thermal stability of oxygen functionalized h-BN and the great variety of
h-BN functionalization options makes functionalized h-BN-s an attractive
platform to design electroactive species for efficient batteries, as proposed
recently by the author of the present work \cite{nemeth2015functionalized}.

\subsection{Charge storage mechanisms in functionalized h-BN}

Charge can be stored in several ways in FBN-s: 

1.) Charge storage in the bond attaching a radical to the surface of h-BN.
The B atom of the h-BN monolayer is electron deficient, its valence shell
can store up to two additional electrons. This becomes clear when one considers the
specific one resonance structure of h-BN in which there are only single ($\sigma$) bonds between
the N and B atoms and the lone pair of N is fully localized on the N p$_{z}$
orbital. In this resonance structure, 
the B valence shell has only six electrons while the N has eight. When a radical
functionalizes the B atom, such as $\cdot$OH or $\cdot$NH$_{2}$, it donates
one electron to the empty B p$_{z}$ orbital which then transforms to an approximate
sp$^{3}$ hybrid orbital participating in a $\sigma$ bond binding to the radical.
This $\sigma$ bond, however, still needs an additional electron to complete itself.
The additional electron may come from the anode during the discharge of the device.
Note that in reality the lack of this latter electron is distributed over all the
local bonding environment of the B atom, creating holes (of electrons) in them and
increasing the conductivity of the FBN as compared to h-BN. This hole creation effect is
greatest if the functionalizing atom has a big electronegativity, such as fluorene, and
indeed this is also experimentally observed by the good electrical conductivity of
fluorinated h-BN \cite{xue2013excellent,du2014one}. Such charge storage mechanism
is justified by the existence of both charge neutral $\cdot$OH radical functionalized
h-BN and OH$^{-}$ anion functionalized h-BN where the large negative charge of the
functionalized h-BN has also been pointed out, see Ref. 
\onlinecite{Fu2016microwave}.

2.) Charge storage in the functionalizing group. It is also possible
to store charge in other parts of the functional group, not only in the bond that
links the functional group to h-BN. This charge storage may be based on the
(reversible) reduction/oxidation of chemical bonds in the functional group,
such as for example epoxy ring opening or C=O to C-O$^{-}$ reduction during
discharge. This charge storage mechanism may involve bond breaking and bond formation.
A mixture of the two kinds of charge storage may also simultaneously occur
as illustrated by the example of the BN(F)-BN(OBF$_{2}$) system discussed in the following.  


\section{Results and Discussion}

Perhaps the most fundamental FBN material is oxygen functionalized h-BN, the
analogue of GO. Early studies focused on the high temperature 
oxidation processes of h-BN and its composites to B$_{2}$O$_{3}$ and boric acide as
well as boron glasses \cite{jacobson1999high1,jacobson1999high2}. Only recent studies
attempted the production of oxygen functionalized nanoplatelets and 
mono/few layers of h-BN, using high temperature (800-1000 $^{o}$C 
heating on air in a sealed quartz tube \cite{li2014strong,cui2014large}).
When moisture is present during the high temperature heating, quick transformation
to B$_{2}$O$_{3}$ and boric acide happens \cite{jacobson1999high1,jacobson1999high2}.
Exposing h-BN to air-plasma results in hydroxyl-functionalized h-BN, instead of oxygen
functionalized one, due to the presence of water in the air \cite{pakdel2014plasma}.
Theoretical studies of ozone absorption on h-BN also indicate the possibility of epoxy  
group formation on its surface \cite{anota2011lda}.
Long exposure of h-BN to heating in air at a moderately high temperature of 600
$^{o}$C results in surface coverage by oxygen species with additional boric oxide
formation \cite{jin2016functionalization}. 

\begin{figure*}[tb!]
\resizebox*{6.4in}{!}{\includegraphics{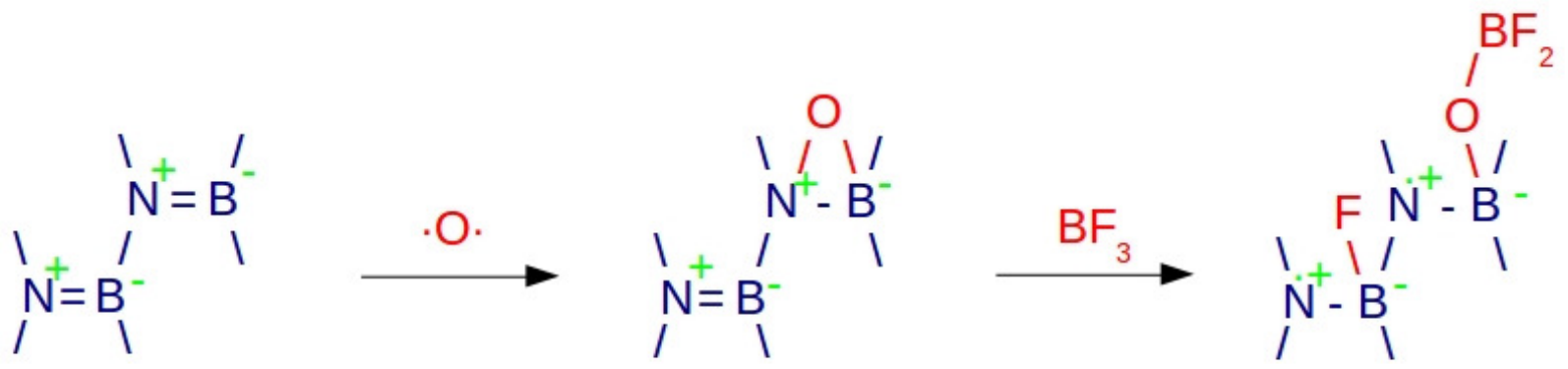}}
\caption{
Proposed synthetic steps toward BN(F)-BN(OBF$_{2}$) on a two-BN-unit fragment of h-BN: 
the epoxy-functionalization of half of
the B=N bonds is followed by the reaction with strong Lewis acid BF$_{3}$. Green indicates
the charge of the B and N atoms of the h-BN monolayer in the given resonance structures, 
``$\cdot+$'' (``dot'' and ``plus'') indicates half of the lone pair in N, i.e.
a hole localized in the N lone pair on the N p$_{z}$ orbital. 
The B=N double bonds in the given resonance structure are due to the donation of the
lone pair of N toward B in a $\pi$-bond.
The charge distribution reflects this resonance structure, in reality the 
N charge is (much) more negative, the
B charge is more positive due to the much greater electronegativity of N as compared to B.
Red indicates the atoms and bonds introduced by functionalization.
The oxygen radical in the first step may come from various sources, such as ozone,
for example. The second step of the process is analogous to that recently seen in the GO
$+$ BF$_{3}$ reaction \cite{samanta2013highly}.
}
\label{ReactionEquation1}
\end{figure*}

Another recent study investigated the fluoroboration reaction of GO with the ether
adduct of BF$_{3}$ \cite{samanta2013highly}. In this reaction the epoxy rings on the
surface of GO open up forming a C-F and a C-OBF$_{2}$ unit on the C atoms of the original
epoxy unit. Such a functionalization of the surface is advantageous for two reasons:
1. Fluorene functionalization may lead to high voltage cells and 2. the -OBF$_{2}$ unit
may capture the F$^{-}$ released during discharge forming -OBF$_{3}^{-}$: a
mechanism to make carbon-fluoride batteries rechargeable as suggested in Ref.
\onlinecite{jones2011polymer}. 
Unfortunately, the C-OBF$_{2}$ unit is not stable,
and -OBF$_{2}$  is generally a good leaving group in the presence of Lewis bases, such as
thiols and amines \cite{samanta2013highly}. As opposed to the C-OBF$_{2}$ bond on the
surface of functionalized graphene, 
the B-OBF$_{2}$ on the surface of functionalized h-BN may be much more 
stable owing to the electron-deficiency of boron in h-BN. Also, the B-O single-bond energy
is much greater (536 kJ/mol) than that of C-O (358 kJ/mol) or N-O (201) 
\cite{cottrell1958strengths,darwent1970bond}. Furthermore, the existence of  
singly negatively charged tetravalent borate ions, such as the
tetrahydroxyborate anion or the related polyborates, points toward the stability of
negatively charged sp$^{3}$ hybridized boron in FBN-s. The quantum-chemical
calculations discussed below support this expectation.
The respective functionalization process of h-BN to BN(F)-BN(OBF$_{2}$) is depicted in 
Fig. \ref{ReactionEquation1}.
The following sections will describe the band structure evolution of h-BN upon
step(layer)-wise functionalization, first by oxygen, then by BF$_{3}$, followed by
cell reactions of the product with Li, Na and Mg anodes, predicting the voltages
and energy densities of these cells. Furthermore, the applicability of these
materials as high voltage pseudocapacitors will also be discussed.

\subsection{Computational methodology}
\begin{figure*}[tb!]
\resizebox*{5.8in}{!}{\includegraphics{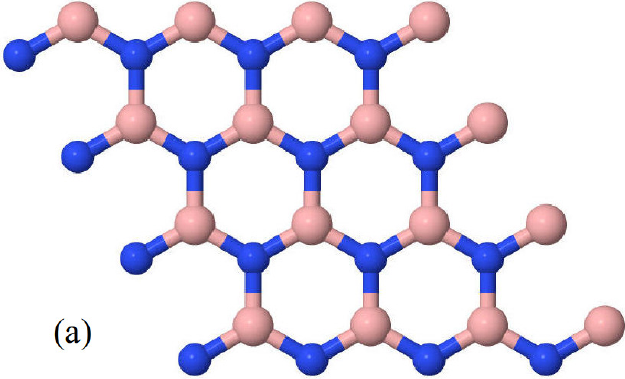}\includegraphics{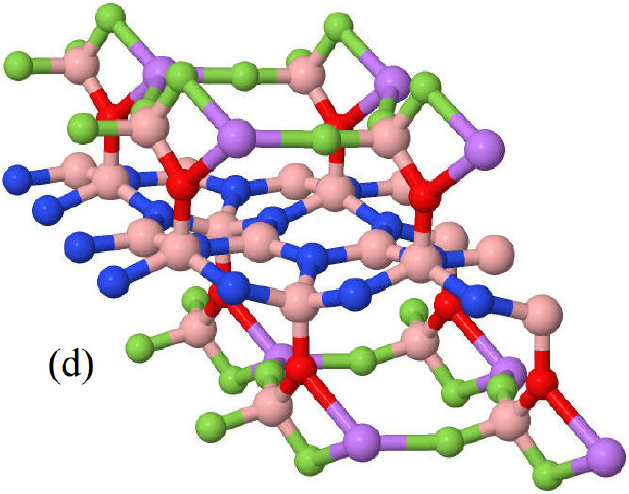}}
\resizebox*{5.8in}{!}{\includegraphics{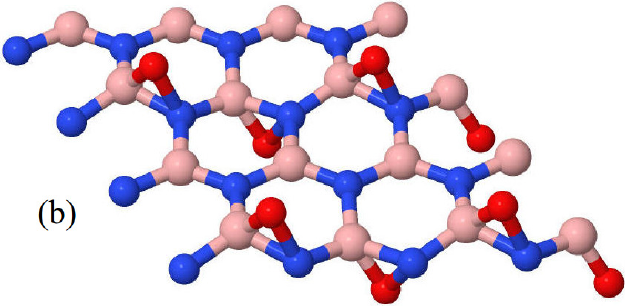}\includegraphics{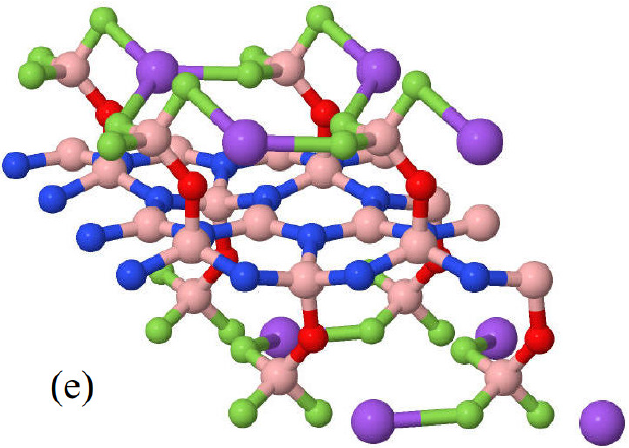}}
\resizebox*{5.8in}{!}{\includegraphics{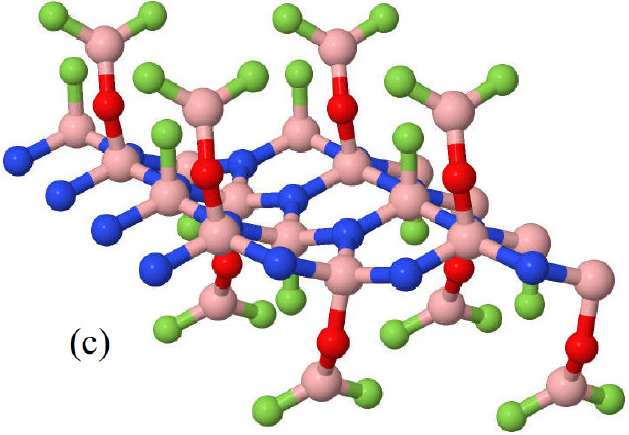}\includegraphics{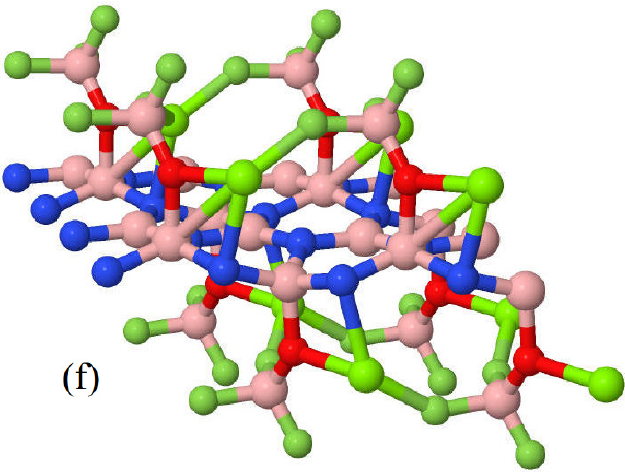}}
\resizebox*{5.8in}{!}{\includegraphics{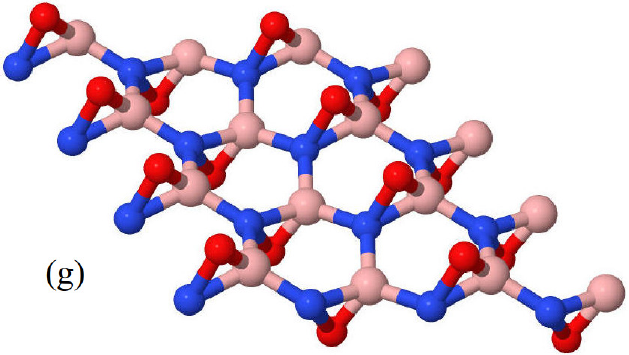}\includegraphics{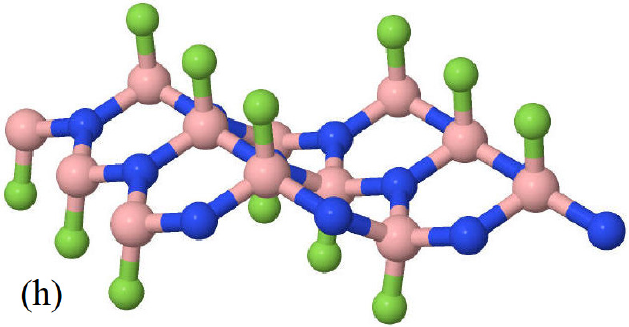}}
\caption{Fragments (supercells) of examples of functionalized h-BN. 
Color code: B - magenta, N - blue, O - red, F - green (small),
Li - violet (small), Na - violet (large), Mg - green (large).
Systems shown: h-BN, (BN)$_{2}$O,
BN(F)-BN(OBF$_{2}$), BN-BN(OBF$_{3}$)Li, BN-BN(OBF$_{3}$)Na,
BN-BN(OBF$_{3}$)Mg, BNO and BNF,
in panels a through h, respectively. The order indicates step(layer)-wise
funtionalization of a h-BN monolayer, 
first by epoxide groups on every 2$^{nd}$ B-N bonds, then the
epoxyde groups are reacted with BF$_{3}$ forming BN(F)-BN(OBF$_{2}$) units, then
discharge products of the latter 
with Li$^{+}$, Na$^{+}$ and Mg$^{2+}$ cations and BN-BN(OBF$_{3}^{n-}$) 
anionic units (n=1 or 2), finally the fully epoxy-functionalized (BNO) and the fully
(on all B atoms) fluorene functionalized h-BN (BNF) is shown for comparison.
}
\label{picsFBNs}
\end{figure*}

DFT calculations have been carried out using the Quantum Espresso program package
\cite{QE-2009,QE-2017}, following the methodology in 
Refs. 
\onlinecite{nemeth2014materials,nemeth2014ultrahigh,zhang2016experimental}. 
A plane wave basis set with 50 Ry
wave-function cut-off has been used along the PBEsol exchange-correlation functional
\cite{PBE,PBEsol} with ultrasoft pseudopotentials as provided by the software package.
Functionalized h-BN monolayers were separated from each other by about 30 {\AA} vacuum
layers and 3D periodic boundary conditions were applied. Each simulation cell
involved a double stoichiometric formula unit of the respective compound in order to allow for
uniform functionalization on both the top and bottom surfaces of h-BN. 
Supercells of the simulation cells are shown in Fig. \ref{picsFBNs}.
Spinpolarized calculations have
been carried out to allow for open shell systems characteristic for radicals-based functionlization
of h-BN. The k-space grid was 10x10x4, this allows for a mRy
convergence of the electronic energy with respect of k-space saturation and geometry
convergence. This accuracy is satisfactory for the prediction of accurate cell
voltages within the chosen exchange-correlation functional and surface model. The
surface models have been relaxed until residual forces on the atoms became smaller
than 0.0001 Ry/bohr and residual pressure on the simulation cell was less than 3
kbar. Energies of cell reactions were calculated from electronic energy differences
of products and starting materials, all in the crystal phase. Cell voltages are
computed as the negative of the cell energies divided by the number of electrons
transferred in the reaction.  The methodology has been validated on experimental
data of lattice parameters and enthalpies of formation of $\alpha$-Li$_{3}$BN$_{2}$, 
Li$_{3}$N, and h-BN. Since the model systems have the same translational symmetry as
h-BN, only with larger unit cells, the representation of the electronic 
band structure used the same high symmetry k-points.
Enthalpies of formations were estimated as the change of  electronic energy during
the formation of the compounds from the corresponding elements at T =  0 K, i.e.,
from crystalline Li and B and N$_{2}$ gas. Experimental lattice parameters have been
reproduced within 2.5\% error, while experimental enthalpies of formation were within
4\% \cite{nemeth2014ultrahigh}. 
The coordinates of the high symmetry points in the k-space are $\Gamma$(0,0,0), M(1/2,0,0) 
and K(2/3,1/3,0).

While the PBEsol functional typically underestimates band gaps, it usually
provides good structural parameters and relative energies, such as in the above
validation cases. Hybrid functionals, such as PBE0 
\cite{PBE01perdew1996rationale,PBE02adamo1999toward} and HSE \cite{heyd2003hybrid}
provide more accurate band gaps than PBEsol, as pointed out in band structure studies on h-BN 
as well \cite{zolyomi2015towards}, however, they are computationally much more expensive
and not fully available in Quantum Espresso yet.
Even if the PBEsol band gaps would underestimate the real ones by about 1 eV, it would
likely not change the relative order of them, i.e. PBEsol is assumed to be accurate
enough to separate large, medium and small band gap systems to estimate what can be
expected for the electrical conductivity of the given systems. PBEsol band gap values
can also be compared to known band gaps of similar FBN-s, such as oxidized,
hydroxilated, fluorinated or fluorosulfonated ones, to check whether the magnitude of
the band gaps have been qualitatively well described. For example, highly electron
withdrawing fluorene or fluorosulfonyl radicals are experimentally known to create
strong p doping that results in small bandgap FBN-s with strong magnetism. The
fluorosulfonyl-analogous -OBF3 functionalization is expected to result in small band
gap as well and PBEsol is capable to verify or at least support this expectation.
Therefore, the accuracy of PBEsol is sufficient to provide reliable guidance for the
selection of target cathode materials for experimental testing, which is the main goal
of the current work.

\subsection{Band structure evolution due to functionalization}
Fig. \ref{ReactionEquation1} depicts the process of the stepwise functionalization of
h-BN. In the first step, oxygen radicals attach to neighboring B and N atoms, forming
epoxy rings on the top or bottom surfaces of the h-BN monolayer. In the second step,
these epoxy rings react with BF$_{3}$ in an analogous fashion as formerly seen for GO
\cite{samanta2013highly}. As a result, pairs of nearest B atoms become covalently functionalized
with -F and -OBF$_{2}$ units. Fig. \ref{picsFBNs} shows the evolution of a h-BN sheet
during the functionalization in panels a-c, assuming all boron atoms of the sheet will be
functionalized at the end leading to BN(F)-BN(OBF$_{2}$). Panels d-f indicate the 
discharge products of a BN(F)-BN(OBF$_{2}$) cathode with Li, Na and Mg anodes.
As indicated by the bond lengths in Table \ref{bonds}, both the charged and
discharged systems consist of strong covalent bonds withing the FBN and the cations
intercalate the functional groups of the surface.
Some of the B-N single bonds become as long as 1.59-1.62 {\AA} due to the
functionalization, however these values are well within observations in related B-N
bond containing systems. For comparison, the B-N bond in h-BN is 1.45 {\AA} long, 
in cubic-BN 1.57 {\AA} (sphalerite type) and 1.55-1.58 {\AA} (wurtzite type),
in ammonia-borane 1.58 {\AA} 
and in H$_{3}$N-BF$_{3}$ or Me$_{3}$N-BF$_{3}$ about 1.66 {\AA} \cite{jonas1994studies}.

\begin{figure*}[tb!]
\resizebox*{6.4in}{!}{\includegraphics{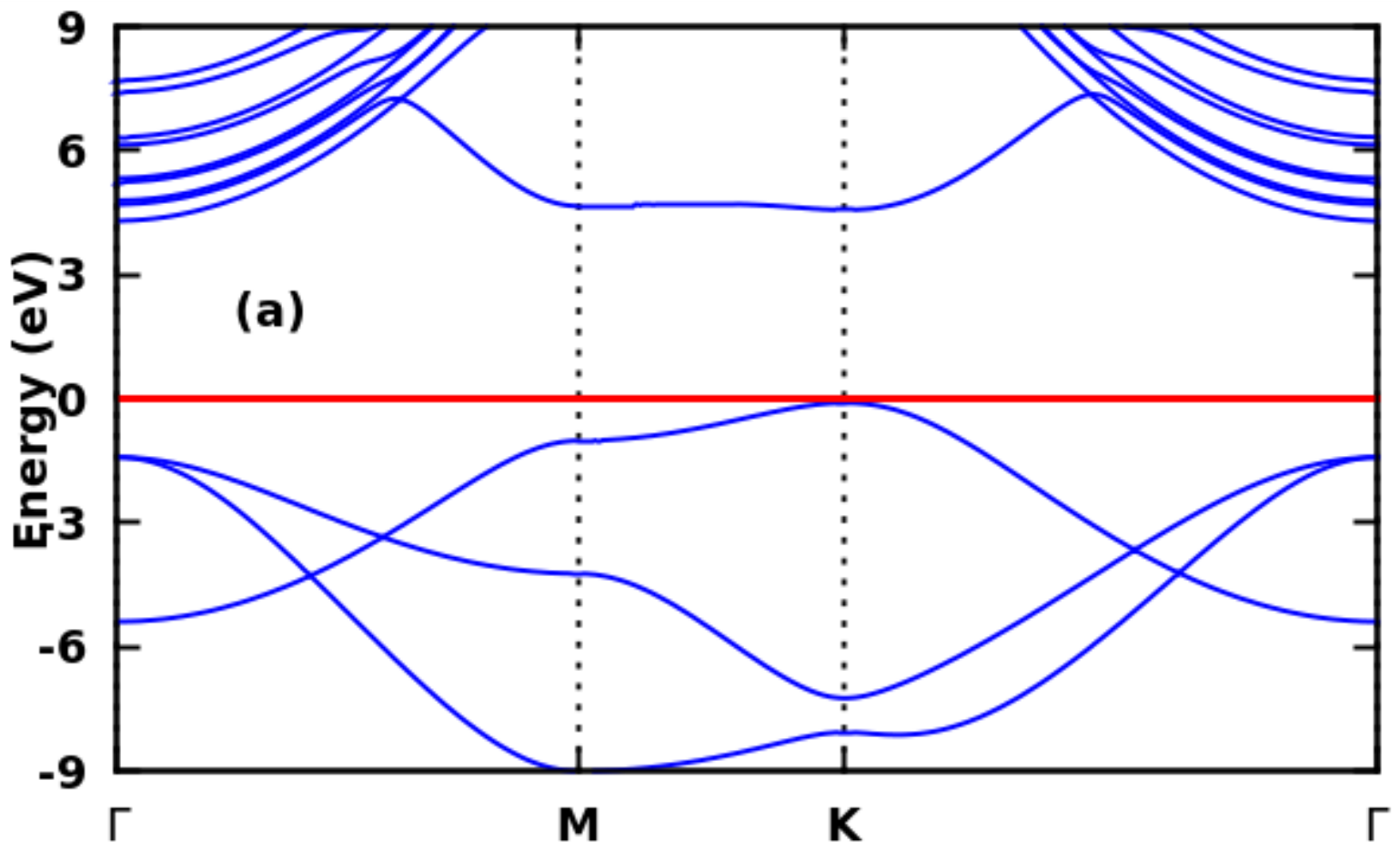}\includegraphics{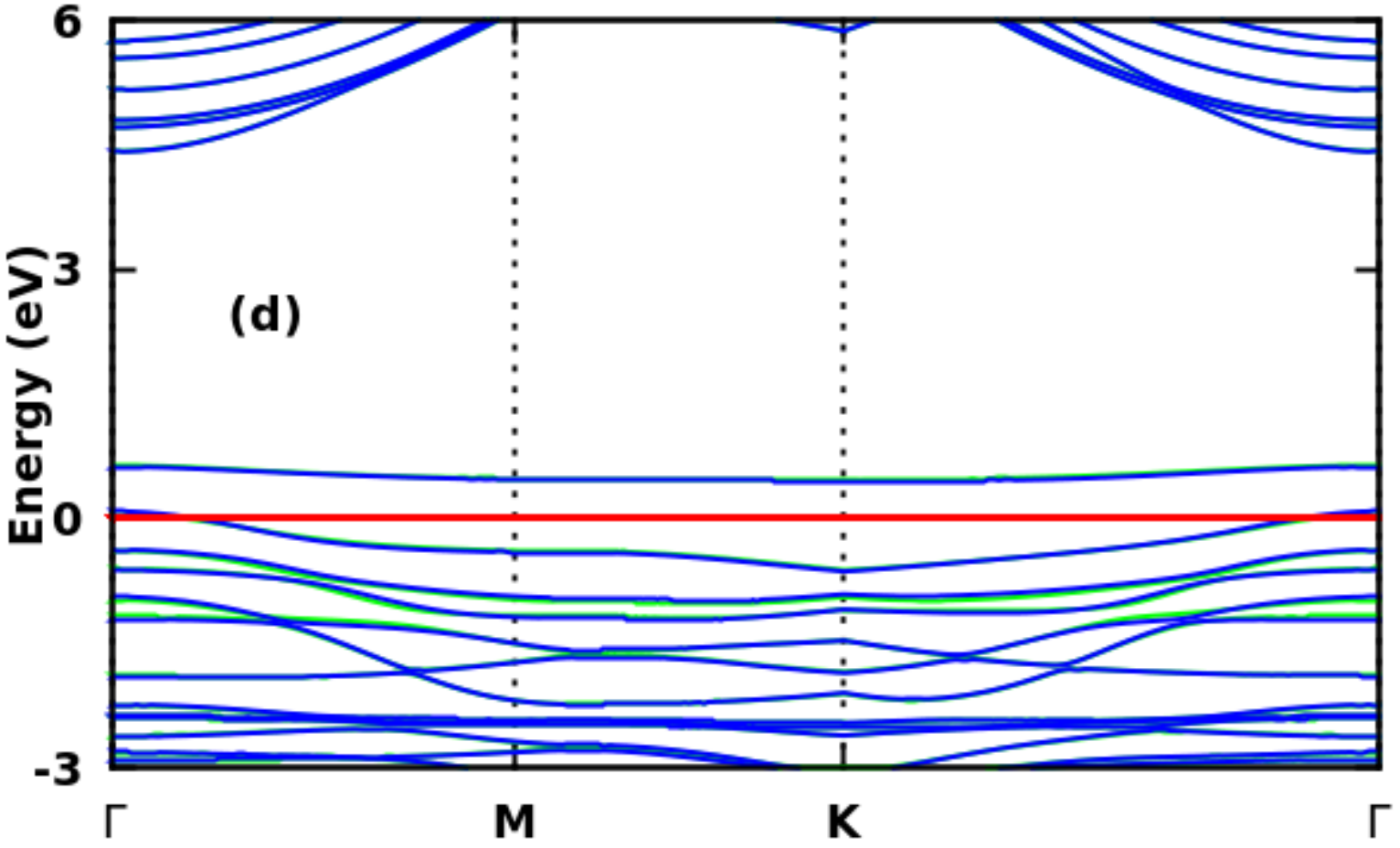}}
\resizebox*{6.4in}{!}{\includegraphics{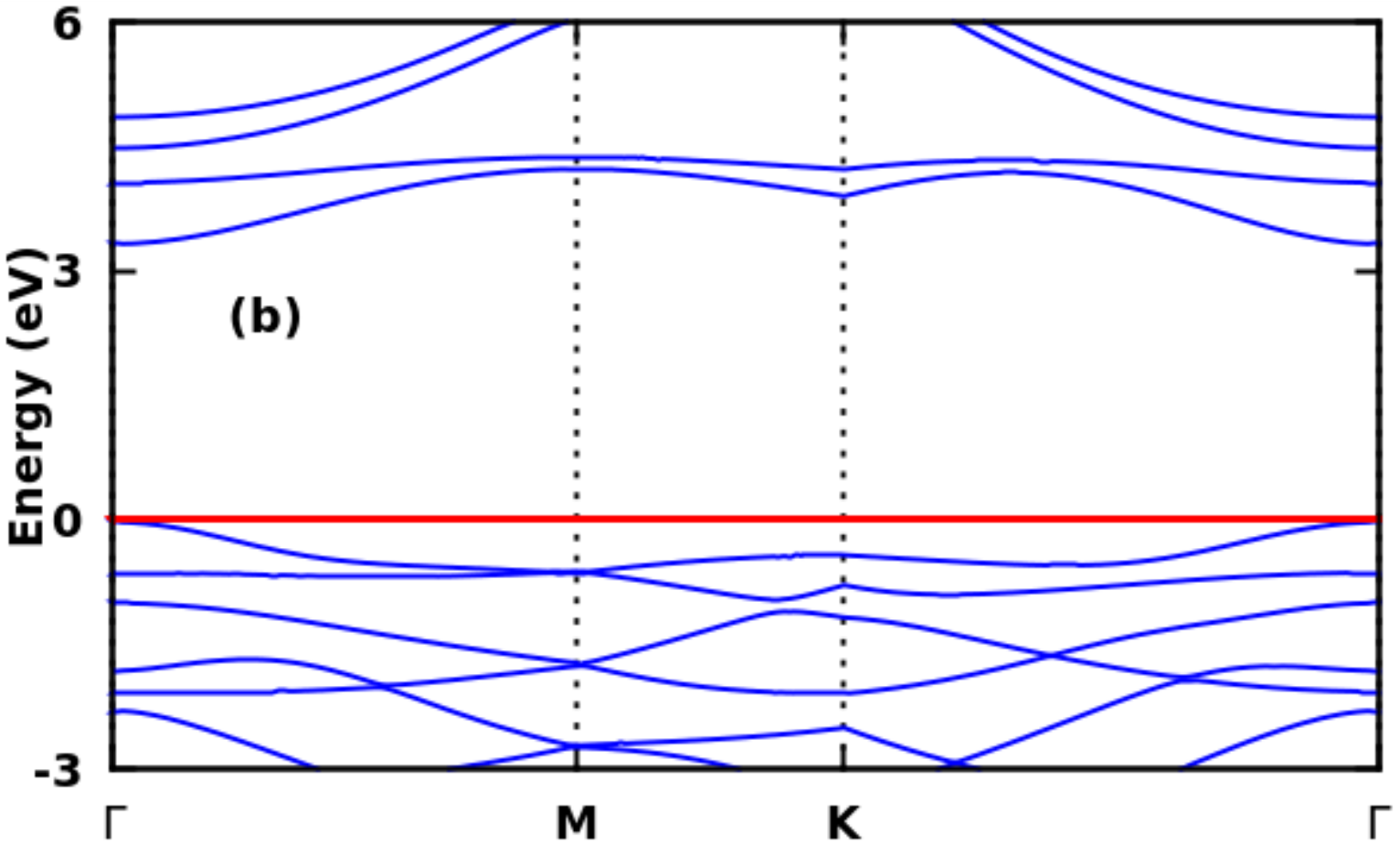}\includegraphics{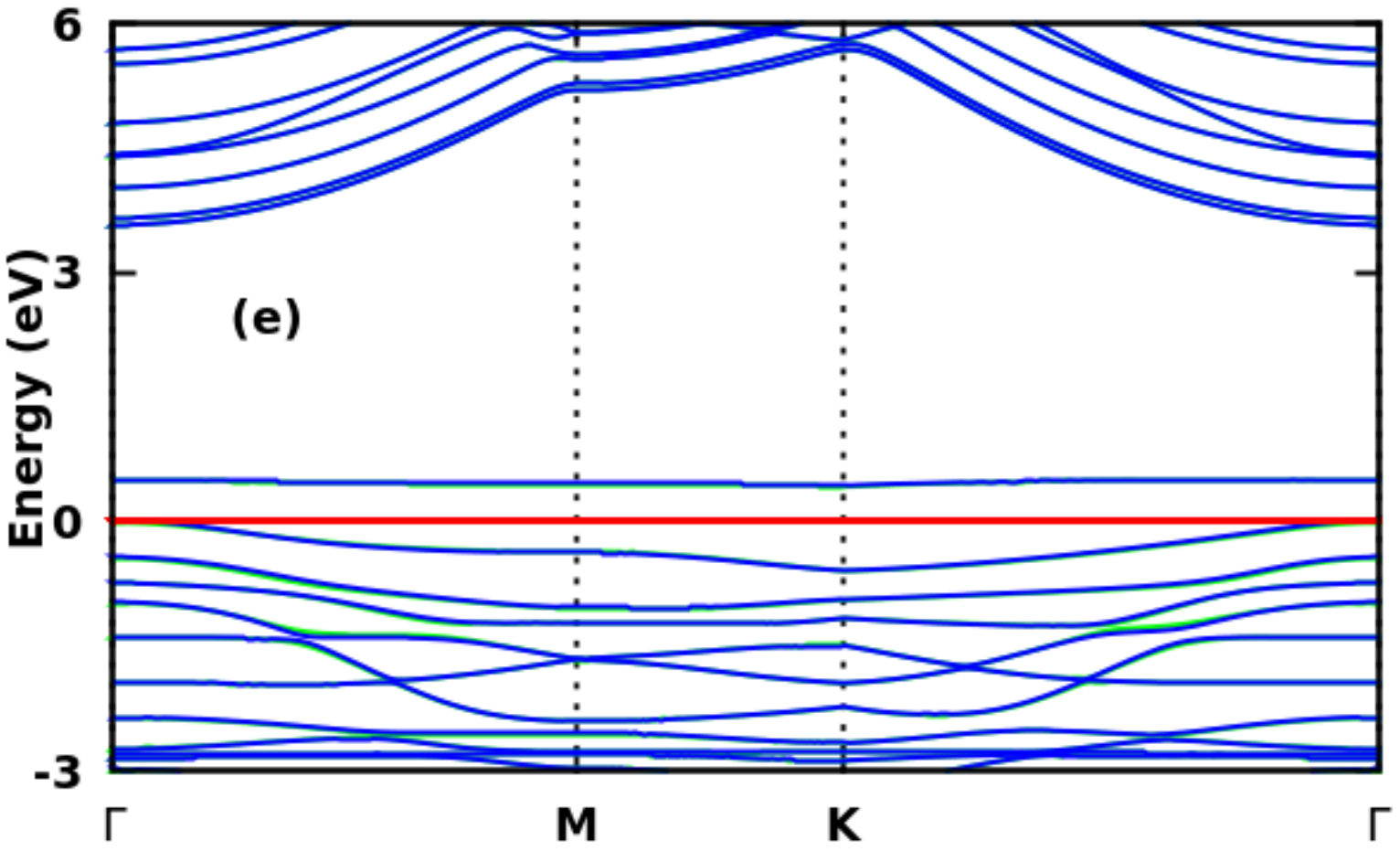}}
\resizebox*{6.4in}{!}{\includegraphics{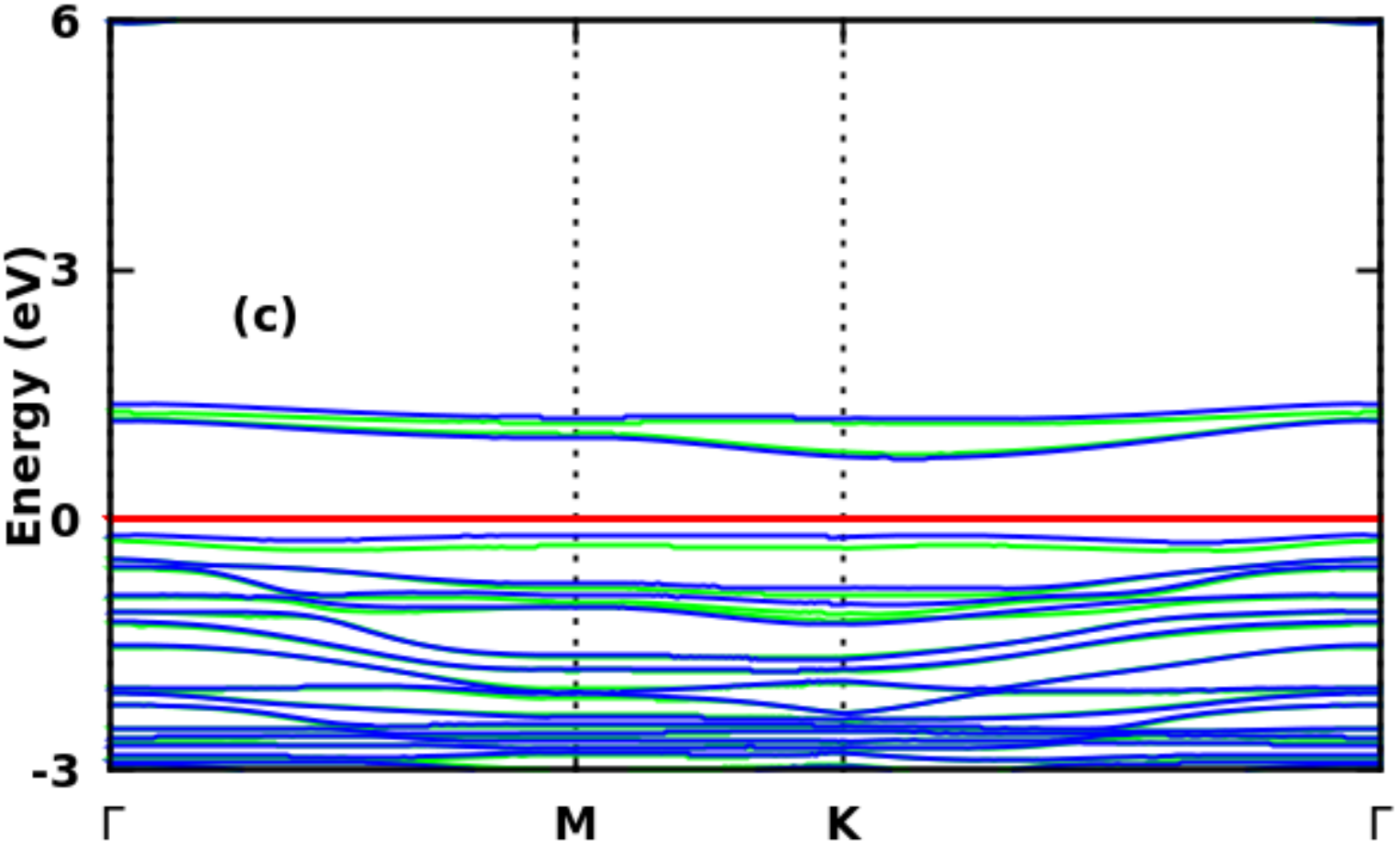}\includegraphics{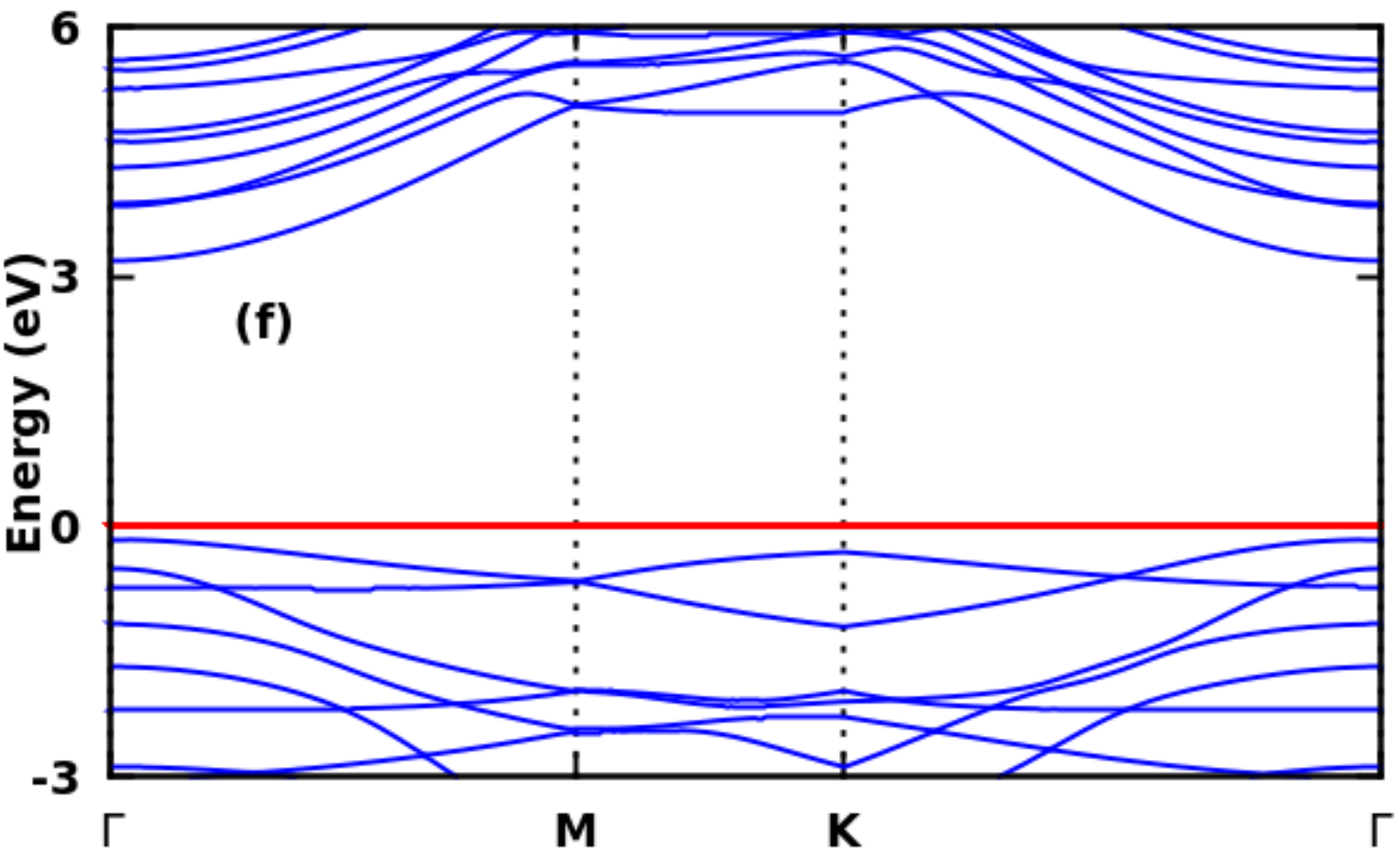}}
\resizebox*{6.4in}{!}{\includegraphics{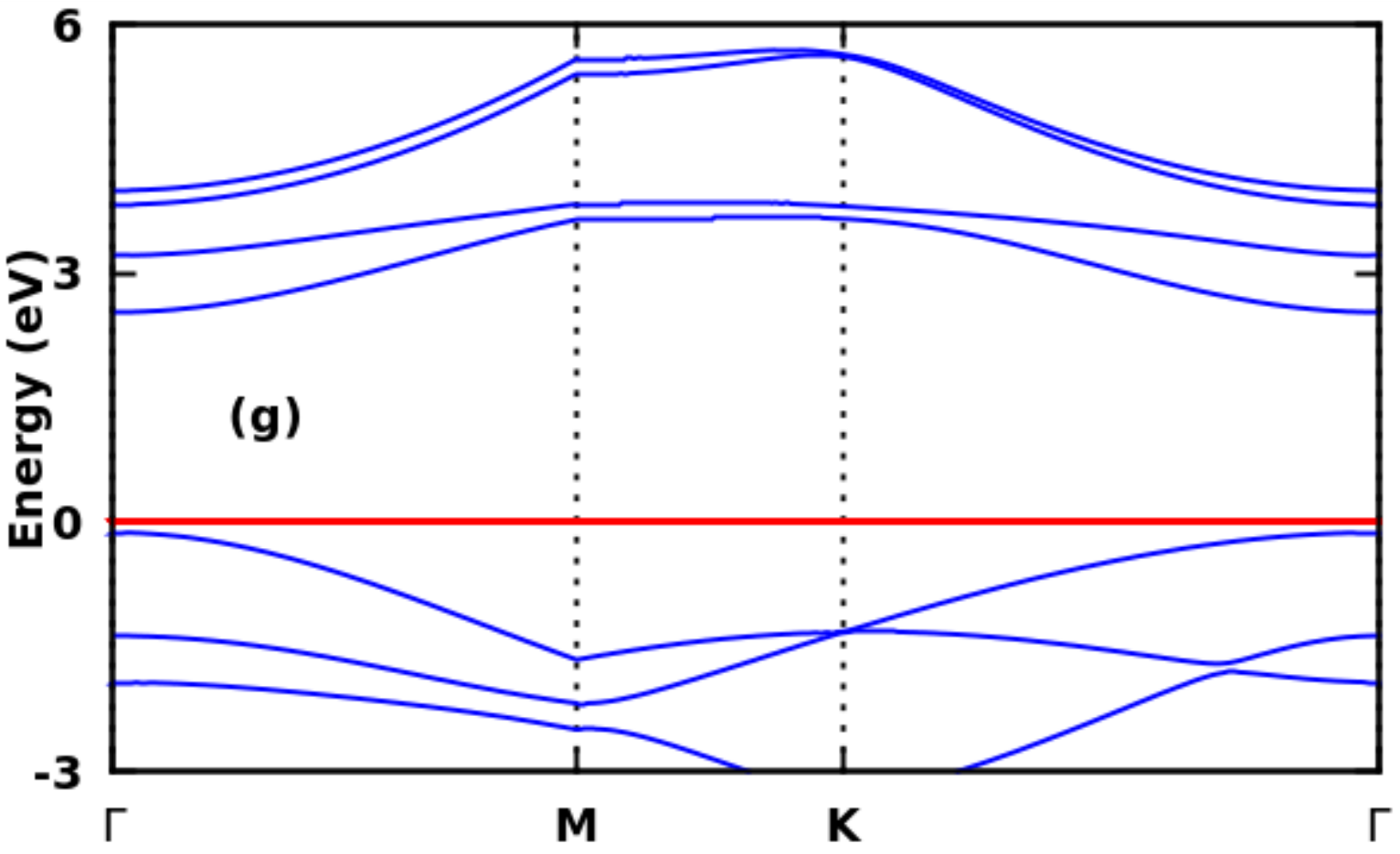}\includegraphics{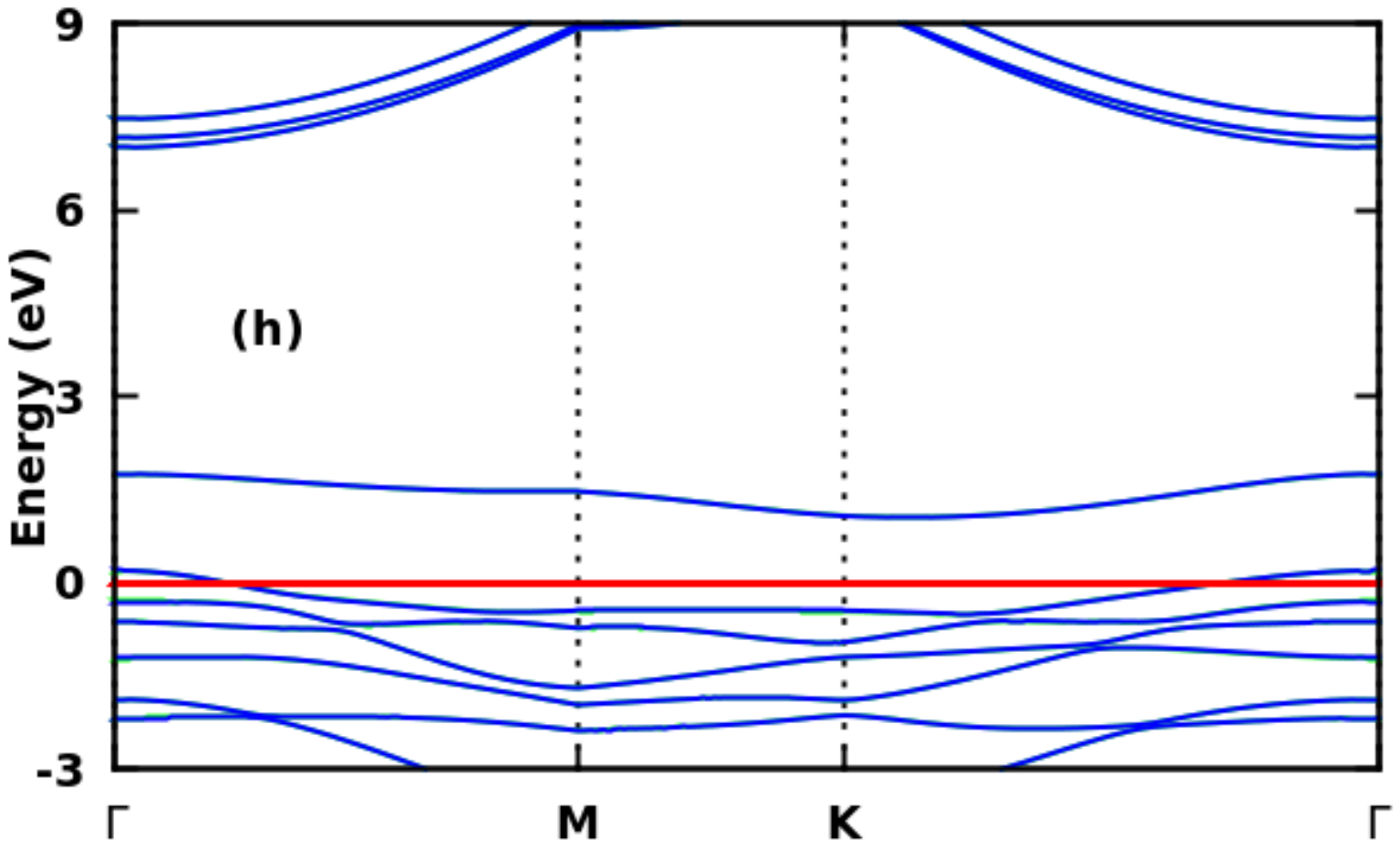}}
\caption{
Spinpolarized calculated band structures of h-BN and its functionalized derivatives. 
Different colors indicate up and down spin bands.
Systems are identical to those shown
in Fig. \ref{picsFBNs} and in the same respective order: h-BN, (BN)$_{2}$O,
BN(F)-BN(OBF$_{2}$), BN-BN(OBF$_{3}$)Li, BN-BN(OBF$_{3}$)Na,
BN-BN(OBF$_{3}$)Mg, BNO and BNF in panels a through h.
The origin of the energy scale is the Fermi energy of the respective system. 
}
\label{bandsFBNsSpinpol}
\end{figure*}

The band structures of the materials discussed in
the present work are depicted in \ref{bandsFBNsSpinpol}.
The respective band gap and magnetization values are presented in Table \ref{bandgaps}.

Pristine (non-functionalized) 
h-BN has an indirect band gap of about 4.4 eV and a direct one of 4.7 eV 
(panel a of Fig. \ref{bandsFBNsSpinpol}), about 1 eV smaller than the experimental band
gap of 5.6 eV.

Oxidation of h-BN reduces the band gap to about 3.3 and 2.7 eV in (BN)$_{2}$O and
BNO, respectively (panels b and g of Fig \ref{bandsFBNsSpinpol}). Note
that in (BN)$_{2}$O and BNO only non-spinpolarized solutions were found. The band-gap of
hydroxilated h-BN has experimentally been found to be 3.9 eV \cite{nayak2014inversion}. Doping
of h-BN with O (substitution of N by O) may lead to an as low as 2.1 eV band gap, according to
experimental observation \cite{weng2017tuning}.

Spin-polarized calculations result in 0.8 eV gap for BN(F)-BN(OBF$_{2}$) and 
zero gap for its alkalinated versions
(panels c, d and e in Fig \ref{bandsFBNsSpinpol}, respectively). The zero bandgaps of the
latter systems appear close to the $\Gamma$-point and the highest energy occupied band crosses
the Fermi energy only to a small extent, therefore also the next smallest gaps are given
in Table \ref{bandgaps}, and these are of 0.6 eV.

Note that the band gaps in BN(F)-BN(OBF$_{2}$) and its Li and Na reduced versions 
are quite small, much smaller than that of some
semiconducting polymers, such as for example polyacetylene, in which the band gap is about
1.4 eV. In the most common Li-ion battery cathode material, LiCoO$_{2}$,
the band gap is 2.7 eV \cite{van1991electronic} and it reduces to zero as the cathode is
charged and its Li content per formula unit decreases below 0.75 
\cite{molenda1989modification}. Also note that the electrical conductivity of
LiCoO$_{2}$ is in the order of 0.0001 S/cm and can be increased to 0.5 S/cm by Mg
doping \cite{park2010review,tukamoto1997electronic}. Fluorosulfonyl functionalized h-BN
has a higher electrical conductivity of 1.5 S/cm \cite{shen1999intercalation}.           

Calculations on the Mg reduced form of BN(F)-BN(OBF$_{2}$) resulted in only non-spinpolarized 
solution with a band gap of about 3.4 eV (panel f of Fig. \ref{bandsFBNsSpinpol}).

The fluorene functionalized h-BN hasa direct zero band gap,
characteristic of metals (panel h of Fig \ref{bandsFBNsSpinpol}).
With the above in mind, one possible way to introduce electrically conducting domains 
in h-BN may be based on
fluorination. Also note that the good electrical conductivity of some fluorinated and
fluorosulfonated h-BN has been
experimentally observed \cite{xue2013excellent,shen1999intercalation}. 

Another way to increase electrical conductivity in BN(F)-BN(OBF$_{2}$) is n-type doping that
happens when Li, Na or Mg reduces BN(F)-BN(OBF$_{2}$) during the discharge of the battery.
While the fully discharged systems may have small bandgaps, as a result of filling up the
lowest energy conduction states, partial filling of these conduction states may result in
better electric conductivity. Also note that the band structures in the respective systems 
substantially change during the discharge of the battery as one B-F bond breaks and another
forms, therefore a simple band-filling approach is insufficient to understand how the
band gap changes as a function of the extent of reduction (discharge).
One formula unit of BN(F)-BN(OBF$_{2}$) can in principle take up a maximum of two electrons,
as two electrons are missing to complete the valence shells of the two B-s that are being
functionalized (either by $\cdot$F or by $\cdot$OBF$_{2}$ radicals) 
per formula unit of BN(F)-BN(OBF$_{2}$).
Thus, in BN(F)-BN(OBF$_{2}$)Li and BN(F)-BN(OBF$_{2}$)Na half of the conduction band
, i.e. half of the holes in the nitrogen valence shells,
remains unfilled, which results in zero bandgaps and leads to
improved electrical conductivity of these systems as compared to the fully discharged
BN(F)-BN(OBF$_{2}$)Mg. Note that full discharge is also possible with two Li or Na per
formula unit, however in the present model the solvation shell for the second alkali cation
could not be efficiently represented in the framework of a monolayer system, 
as it should probably be represented via the interlayer interactions that are not discussed
here. Also note that small bandgap is in principle also possible with Mg anode
when only a partial discharge is applied.

Even though BN(F)-BN(OBF$_{2}$) and its (partially) reduced versions may have small bandgaps, 
addition of small amounts of 
graphene may make these systems electrically conductive to the desired degree, as it is
customary with most battery cathode materials, while such graphene composites
would also be ionically conductive without the addition of liquid electrolytes, based on the
ability of Li and Na ions to hop between the surface sites of BN(F)-BN(OBF$_{2}$). 
Thus these sytems are good candidates for cathodes in all-solid-state batteries that 
are highly desirable as especially safe and high energy density batteries. 

Ionic conductivity is assumed to be good in all FBN-s due to the strong
dipoles in the B-N bonds and in the functional groups on the surface, as well as
sufficient space on the surface of the monolayers for virtually all kinds of
ions to move. In fact FBN-s
have been proposed and experimentally tested to some extent 
for use as ionic conductors \cite{mofakhami2015material,kumar2010solid}.

Magnetization values support that the combined oxygen and BF$_{3}$ functionalization creates
holes in all four nitrogen atoms of the simulation cell of 
BN(F)-BN(OBF$_{2}$) (two formula units used per simulation cell for top and bottom surface
functionalizations), as significant magnetization values of
0.30-0.45 $\mu_{B}$ have been calculated only for the N-sites, 
while magnetization values on other atoms are at least an order of magnitude smaller. 
After the reaction with Li or Na, a reduction and rearrangement
to BN-BN(OBF$_{3}^{-}$) ions happens. In these latter anions some N-s will have two B neighbors
with OBF$_{3}^{-}$ functionalization and one non-functionalized B, while some other N atoms
will have two non-functionalized B neigbors and one OBF$_{3}^{-}$ functionalized one. This is
also reflected in the magnetization changes: the N-s with two functionalized B neighbors remain
significantly magnetized,, while the other N-s 
essentially lose their magnetization.
In the case of Mg anodes, two negative charge will be transferred to a formula unit of the
cathode and doubly negative BN-BN(OBF$_{3}^{2-}$) ions form. This results in a total filling of
all holes in the N-s, a complete loss of magnetization and a large band gap of 3.4 eV.

A similar analysis holds to BNF, where all high magnetizations of about 0.45 $\mu_{B}$
happen on the N atoms, while only a tenth of this value occurs on the F-s.
Fluorene or combined oxygen and BF$_{3}$ functionalization represents a p-type doping of h-BN and
is a powerful means to engineer conductivity
and electrode reactivity / rechargeablity of functionalized h-BN. The holes created by these
p-type dopings will be filled when the respective batteries discharge and the discharge of the
battery can be viewed as n-type doping by alkaline or alkaline earth atoms. Thus h-BN provides
a platform that is suitable for p-type doping and recombination of electrons and holes through
a subsequent n-type doping that follows during the discharge of the battery.

Also note that total magnetization values (sum of up and down magnetizations) are near zero in
all magnetic systems mentioned above. This implies an antiferromagnetic ordering of the
individual atomic magnetizations.

Therefore, the conclusion is that it is the presence of holes
in the N atoms of BN(F)-BN(OBF$_{2}$) and its Li, Na and Mg doped (reduced) versions
that is responsible for small band-gaps and magnetizations. The electrical conductivity of these
systems is a function of hole-concentration: too high hole concentration in BN(F)-BN(OBF$_{2}$)
is not optimal but partially n-type doped (reduced) BN(F)-BN(OBF$_{2}$) may lead to 
zero bandgap and good electrical conductivity.

\subsection{Electrochemical features}
During the discharge process, the triangle of the epoxy bond of BNO opens up in the N-O bond
while it stays intact in the B-O bond and converts to a singly
negatively charged BN$_{3}$O$^{-}$ ion that is approximately tetrahedrally shaped and
is linked to the rest of the 2D sheet through its N-corners. 
During discharge, the BN(F)-BN(OBF$_{2}$) cathode material detaches a 
F$^{-}$ ion from its BN(F) unit and captures it on a nearby OBF$_{2}$ unit according
to the intramolecular surface conversion process of
\begin{equation} \label{OBF3reaction}
BN(F)-BN(OBF_{2}) + e^{-} \rightarrow BN-BN(OBF_{3}^{-}) ,
\end{equation}
thereby forming a BF$_{4}^{-}$-like ion, OBF$_{3}^{-}$ that is covalently bound to the h-BN
surface through its O-atom. The resulting discharged 2D monolayer
acts as a 2D array of linked anions, a 2D polyanion. In fact, such a charged 2D
polyanion has been observed in OH$^{-}$ functionalized h-BN and its large negative
charge (zeta-potential) has been measured \cite{Fu2016microwave}.
The discharge product negative ions are charge-balanced by
nearby cations, such as depicted in Figs. \ref{picsFBNs} d-f. 
Note that this mechanism enables the making of a rechargeble battery from a
fluorinated 2D FBN material (BN(F)-BN(OBF$_{2}$)), as opposed to the well known and commercially
available fluorinated graphite cathode material, CF$_{x}$, 
that is not rechargeable due to
irreversible formation of fluoride salts. Also note that a similar mechanism to make
CF$_{x}$ rechargeable was proposed in Ref. 
\onlinecite{jones2011polymer}, 
though it
was based on an inter-, instead of an intra-molecular F$^{-}$ transfer process.
Simple fluorination of h-BN that leads to BNF would be insufficient to mantain rechargeablity
for the same reason CF$_{x}$ is not rechargeable. It is only the more sophisticated
BN(F)-BN(OBF$_{2}$) functionalization that allows for rechargeability.

As summarized in Table \ref{OCV-GED},
the batteries based on BNO are of relatively small open circuit voltage 
($\approx$ 2 V), though this is approximately the voltage of intensely researched 
Li-sulfur batteries, while the ones based on BN(F)-BN(OBF$_{2}$) are of high voltage (3.6-5.6
V) as these latter ones represent reactions between bouound fluorene radicals and Li, Na or Mg. 
Both types of batteries have high theoretical gravimetric energy densities and
capacities. 
 
\subsection{Pseudocapacitive applications}
Two fundamental features of pseudocapacitors involve 1. a ``sloping'' electrode
potential as a function of charge withdrawn or added to the electrode (classically
associated with capacitors, not with batteries where the potential should be approximately
constant during discharge) and 2. chemical (redox)
reactions happening during the charging/discharging process on the surface or in the
bulk of the electrode (classically associated with batteries, not with capacitors) 
\cite{conway1991transition,dubal2015hybrid,conway1999electrochemical}.

2D functionalized h-BN-s can, at least to some extent, contribute to
pseudocapacitive effects. As it is seen from the above described examples of BNO and
BN(F)-BN(OBF$_{2}$), chemical reactions may happen on the surface of these materials,
such as epoxy ring opening/closing or fluorene reduction/oxidation accompanied with 
fluoride ion capture in the OBF$_{2}$ side-chain. Since the negative charges will be
localized in the monolayer surface, all positive ions should also fit in the 
surfaces, such as indicated in Fig. \ref{picsFBNs}, in order to have a battery. 
Positive ions, however, may not always all fit in the surface, for several reasons. 
The primary such reason is the size of these ions. If their solvation shells, or the
ions themselves are large enough, they cannot be packed as dense on the surface of the
monolayers as the negative ions and therefore the positive ions must form multiple
surface layers resulting in a sloping electrostatic potential. To maintain such a
potential, the negative ion side-chains must be bound by strong covalent bonds in
the surface. It is expected that pseudocapacitive behaviour of FBN-s increases 
with the increasing size of the cations used. For example, large organic cations, such
as quaternary ammonium ions are expected to provide large pseudocapacitance with
FBN-s. 

The use of large organic cations may allow for a rechargeable use of BNF (h-BN
fluorinated on all B atoms) as a cathode material in supercapacitors and pseudocapacitors. 
With Li, Na or Mg cations, BNF
would decompose to the respective fluoride salts and h-BN, similarly to CF$_{x}$, as
the close vicinity of these small cations would polarize the B-F bonds too much, the
cations would rip the F$^{-}$ ions off from the h-BN surface and form fluoride salts, 
such as in the BN(F)
unit in BN(F)-BN(OBF$_{2}$) in reaction Eq. \ref{OBF3reaction}. While 
the OBF$_{2}$ side-chains capture the detached F$^{-}$ ions in BN(F)-BN(OBF$_{2}$),
there is no such option in BNF. It is however reasonable to assume that 
the F$^{-}$ ions would not be detached from the h-BN surface with sufficiently large
organic cations, both for the
smaller polarizing effect of these cations and for the thermodynamically less energetic
formation of the corresponding organic fluorene salts (or ionic liquids) as compared
to Li/Na/Mg-fluorides. While BNF would act as the electron withdrawing component of an
organic charge transfer complex, no such behaviour is known for CF$_{x}$. The reason is 
that BNF has holes in electronic orbitals of high electronegativity N atoms,
generated by electron withdrawing effect of nearby B-F bonds (more precisely B-bound F
radicals), while CF$_{x}$ has no such holes.

While most pseudocapacitors are known to operate at small maximum voltages 
(typically $<$ 1 V), some FBN-s may allow for pseudocapacitors with unusually large 
voltages, even as high as about 3 V, when using cathodes like BN(F)-BN(OBF$_{2}$),
owing to the large energy of fluorene reduction/oxidation in the FBN electrode.
As the amount of energy stored in a pseudocapacitor is linearly proportional to its
capacitance and quadratically to its voltage, the applications of FBN
pseudocapacitors may open the way of storing more energy per weight than FBN batteries.

\section{Summary and Conclusions}
The present work proposes new functionalized 2D materials,derivatives of exfoliated 
h-BN as positive electrode electroactive materials for use in 
improved electrochemical energy storage devices.
The choice of
functionalization can tune the electrochemical properties of these materials. Three
such materials have been analyzed in detail: BNO, BNF and BN(F)-BN(OBF$_{2}$). 

These functionalizations represent p-type doping of h-BN that creates holes in the nitrogen valence
shells of h-BN. These holes recombine with electrons during the discharge of the respective
batteries whereby the discharge represents a subsequent, and in the case of BNO and
BN(F)-BN(OBF$_{2}$), reversible n-type doping through the reaction with
Li, Na or Mg atoms.            

BNO and BN(F)-BN(OBF$_{2}$) go through reversible surface conversion reactions during the cycling of
the electrochemical devices and have potential for simultaneously achieving 
high energy and power densities. BNO, BNF and BN(F)-BN(OBF$_{2}$) can potentially also be used 
in pseudocapacitors due to the
dense packing of covalently bound negatively charged functional groups 
in the monolayer surfaces. 
The careful choice of doping levels may lead to good electrical conductivity of these systems
as well.
Since such functionalized h-BN-s are also good ionic conductors, they could form a basis for
lightweight all-solid-state batteries with high energy densities.
\section*{Acknowledgements}

The author thanks Prof. Leon L. Shaw (IIT) 
for helpful discussions and NERSC (U.S. DOE DE-AC02-05CH11231) for
the use of computational resources. 



\begin{table*}[hp]
\caption{
Calculated band gaps (E$_{g}$), total and absolute magnetizations 
for functionalized h-BN-s. 
For BN-BN(OBF$_{3}$)Li and BN-BN(OBF$_{3}$)Na
the band gaps are zero close to the $\Gamma$-point where the highest energy occupied band
crosses the Fermi-energy to a small extent, for more information also the next band gap value
is presented in parenthesis.
}
\label{bandgaps}
\begin{tabular}{lcccc}
\hline   
           &   \multicolumn{3}{c}{spin-polarized} \\
material   &   E$_{g}$ (eV)     & \multicolumn{2}{c}{magnetization ($\mu_{B}$)} \\
                           &                & total & absolute    \\
\hline                                                                     
 h-BN                      &  4.4 (K$\to\Gamma$) / 4.7 (K$\to$K) &  0.0      &   0.0    \\
 (BN)$_{2}$O               &  3.3           &  0.0      &   0.0    \\
 BN(F)-BN(OBF$_{2}$)       &  0.8           &  0.13     &   3.28   \\
 BN-BN(OBF$_{3}$)Li        &  0.0 (0.6)     &  0.01     &   1.53   \\
 BN-BN(OBF$_{3}$)Na        &  0.0 (0.6)     &  0.02     &   1.50   \\
 BN-BN(OBF$_{3}$)Mg        &  3.4           &  0.0      &   0.0    \\
 BNO                       &  2.7           &  0.0      &   0.0    \\
 BNF                       &  0.0           &  0.0      &   1.74   \\
\hline 
\end{tabular}
\vspace{15 cm}
\end{table*}

\begin{table*}[hp]
\caption{
Calculated characteristic bond lengths ({\AA}) at the optimum geometries in the 
cathode materials BNO and BN(F)-BN(OBF$_{2}$), as well as those in BNF, for
comparison. Note that BNF is not stable as a cathode material with Li, Na and Mg
anodes, as the discharge  would result in the respective fluorene salts. 
The anode materials were Li, Na and Mg, the charged systems's anode is referred to as
"charged". 
Mg anode has been used only with BN(F)-BN(OBF$_{2}$).
For comparison, the B-N bond in h-BN it is 1.45 {\AA} long, in ammonia-borane 1.58 {\AA} 
and in Me$_{3}$N-BF3 about 1.66 {\AA} \cite{jonas1994studies}. Also note that the B-F
bonds longer than 1.38 {\AA} are due to the vicinity of a cation.
}
\label{bonds}
\begin{tabular}{lcccccc}
\hline   
Cathode material     & \multicolumn{2}{c}{B-N} & \multicolumn{2}{c}{B-O} & \multicolumn{2}{c}{B-F}\\
                     &charged   &Li,Na,(Mg)    &charged & Li,Na,(Mg)     & charged    & Li,Na,(Mg) \\ 
\hline                                                                                     
 BNO                 &1.57-1.60 &1.55-1.62     &  1.42  & 1.44-1.45      &  -         & -         \\
 BN(F)-BN(OBF$_{2}$) &1.56-1.59 &1.48-1.59     &  1.45  & 1.43-1.46      & 1.32, 1.42 &1.32-1.48  \\
 BNF                 &1.55-1.57 &    -         &   -    &     -          & 1.38       & -         \\
\hline 
\end{tabular}
\vspace{15 cm}
\end{table*}


\begin{table*}[tb!]
\caption{
Calculated open circuit voltages (OCV), 
gravimetric energy (GED) and capacity (GC) densities with two functionalized h-BN
cathode materials: 1. fully epoxy-functionalized h-BN, BNO,
and 2. fluoroborated half-epoxy-functionalized h-BN, 
BN(F)-BN(OBF$_{2}$). The anode materials are bulk metallic Li, Na and Mg.
The corresponding cell reactions involve one electrons for Li and Na and two for Mg per
formula unit of the cathode material.
}
\label{OCV-GED}
\begin{tabular}{lccccccccc}
\hline   
Cathode material     & \multicolumn{3}{c}{OCV (eV)} & \multicolumn{3}{c}{GED (Wh/kg)} & \multicolumn{3}{c}{GC (mAh/g)}\\
                     &   Li     &  Na    & Mg   & Li     &  Na    &  Mg  & Li     & Na    &  Mg \\ 
\hline  
 BNO                 &  2.15    & 2.10   & -    & 1209   & 868    & -    & 561    & 413   & -   \\
 BN(F)-BN(OBF$_{2}$) &  5.60    & 5.10   & 3.6  & 1067   & 874    & 1223 & 191    & 171   & 340 \\
\hline 
\end{tabular}
\vspace{15 cm}
\end{table*}

\clearpage

\end{document}